\documentclass[english,iop]{emulateapj}
\usepackage[T1]{fontenc}
\usepackage[latin9]{inputenc}
\usepackage{textcomp}
\usepackage{amstext}
\usepackage{graphicx}
\usepackage{babel}

\usepackage[dvips]{color}
\usepackage[normalem]{ulem}

\begin{document}

\title{The Thomson Surface: II. Polarization}

\author{C.~E. DeForest$^{1}$, T.~A. Howard$^{1}$, and S.~J. Tappin$^{2}$}

\affil{$^{1}$Southwest Research institute, 1050 Walnut Street Suite 300,
Boulder, CO 80302}

\affil{$^{2}$National Solar Observatory, Sunspot, NM 88349}

\email{deforest@boulder.swri.edu}
\begin{abstract}
The solar corona and heliosphere are visible via sunlight that is
Thomson-scattered off of free electrons, yielding a radiance against
the celestial sphere. In this second part of a three-article series,
we discuss linear polarization of this scattered light parallel and
perpendicular to the plane of scatter in the context of heliopheric
imaging far from the Sun. The difference between these two radiances,
(\emph{pB}), varies quite differently with scattering angle, compared
to the sum that would be detected in unpolarized light (\emph{B}). The
difference between these two quantities has long been used in a
coronagraphic context for background subtraction and to extract some
three-dimensional information about the corona; we explore how these
effects differ in the wider-field heliospheric imaging case where
small-angle approximations do not apply. We develop an
appropriately-simplified theory of polarized Thomson scattering in the
heliosphere, discuss signal-to-noise considerations, invert the
scattering equations analytically to solve the three dimensional
object location problem for small objects, discuss exploiting
polarization for background subtraction, and generate simple forward
models of several classes of heliospheric feature.  We conclude that
\emph{pB} measurements of heliospheric material are much more
localized to the Thomson surface than are \emph{B} measurements, that
the ratio \emph{pB/B} can be used to track solar wind features in
three dimensions for scientific and space weather applications better
in the heliosphere than corona; and that, by providing an independent
measurement of background signal, \emph{pB} measurements may be used
to reduce the effect of background radiances including the stably
polarized zodiacal light.
\end{abstract}


\section{\label{sec:Introduction}Introduction}

Coronagraphs and heliospheric imagers observe sunlight that has been
Thomson scattered off free electrons in the corona and solar wind.
This potential was realized with the invention of the coronagraph
\citep{Lyot1939}, and solar wind transients such as coronal mass
ejections (CMEs) have been observed with ground-based coronagraphs
since the 1950s \citep[e.g.][]{DeMastus1973}. These were accompanied
by spacecraft coronagraphs in the 1970s \citep[e.g.][]{Koomen1975,MacQueen1974}
and the coronagraph legacy continues to this day.

The physics by which this light is scattered are well established,
with the original theory predating the discovery of the electron \citep{Schuster1879}.
Other important developments include the work of \citet{Minnaert1930}
and \citet{Billings1966}. The latter is the publication most commonly
referred to when discussing Thomson scattering theory with regard
to white light observations. The utility of this theory to identify
physical properties (such as mass) in solar wind transient phenomena
(such as CMEs) observed by coronagraphs is well known. Early works
include \citet{Gosling1975,Hildner1975,Rust1979}; and \citet{Webb1980}. 

In the last decade, coronagraphs have been accompanied by another
type of white light observer. Heliospheric imagers, first SMEI \citep{Eyles2003}
and then the \emph{STEREO/}HIs \citep{Eyles2009}, observe Thomson
scattered light at much larger angles ($>20^{\circ}$) from the Sun,
and their ability to track transients such as CMEs has been demonstrated
\citep[e.g.][]{Tappin2004,Howard2006,Webb2006,Harrison2008,Davis2009}.
All current heliospheric imagers observe unpolarized Thomson scattered
sunlight, in part because only recently \citep{DeForest2011} has it
become clear that the bright stellar background could be subtracted with
sufficient precision for quantitative analysis of the faint Thomson-scattered
signal from an imaging instrument.

Although heliospheric imagers use the same scattering physics as coronagraphs,
the wide viewing angle leads to significantly different geometry and
requires different treatment. For example, Thomson scattering becomes
much simpler in the heliospheric case because the Sun can be treated
as a near-point source, eliminating the need to carry van de Hulst
coefficients \citep{Minnaert1930,vandehulst1950} when performing
scattering/radiance calculations. More immediately, coronagraphs are
often assumed to operate near the sky plane, but the relevant figure
for a wide-field imager is the ``Thomson surface'' defined by the
locus of the point of closest approach to the Sun of each line of
sight from the observer. That locus is the sphere with diameter passing
between the Sun and the observer \citep{Vourlidas2006}. Paper I of
this series \citep{Howard2012} covered the applied theory of Thomson
scattering to describe the relationship between this broad-field geometry,
illumination, and scattering efficiency in unpolarized heliospheric
imaging, and demonstrates that a fortuitous cancellation yields a
broad plateau (the ``Thomson Plateau'') of nearly uniform radiance
sensitivity to electron density. 

In the present paper, II of a planned series of three, we explore the
consequences of Thomson scattering theory for polarized light in the
heliospheric context, and discuss scientific applications of the
theory. In particular, we invert the scattering equations analytically
to show how polarized Thomson scattering imagery can be used to
determine the three-dimensional location of individual small
heliospheric features, without the front/back ambiguity present in
similar efforts with coronagraphs \citep[e.g.][]{Dere2005}, and
explore analytically the limits of the technique.  Further, we
demonstrate via a simple forward model that the polarization signal
remains present even for large features that whose position cannot be
solved for analytically.  We also discuss the stability of the
polarization signal from the Zodiacal light and its implications for
measuring the absolute radiance, rather than merely feature-excess
radiance, of Thomson scattered light from heliospheric electrons.

Polarization measurement of Thomson scattered light observation has
existed since the dawn of the coronagraph \citep{Lyot1933} and has
been used for three dimensional analysis of CMEs since shortly after
their discovery \citep[e.g.][]{Poland1976,Wagner1982,Crifo1983}. The
\emph{Skylab} coronagraph, \emph{Solwind}, C/P on board \emph{SMM,}
and LASCO on board \emph{SOHO} all had polarizing
capabilities. Perhaps because polarized coronagraph imagery, requiring
photometry, is harder to work with than is unpolarized imagery
\citep[e.g.][]{MacQueen1993}, it has not been fully exploited in the
spaceflight context although recent work
\citep{Dere2005,Moran2010,deKoning2011} may indicate a renaissance of
polarized image exploitation in the corona.

Polarized detection of Thomson scattered light from CMEs at wide
angles from the Sun dates back much farther than direct
heliospheric imaging cameras such as SMEI and
\emph{STEREO}/HI.  The Helios spacecraft photometers
\citep[e.g.][]{Leinert1975} were used both to characterize the
zodiacal light \citep[e.g.][]{Leinert1981,Leinert1989} and also to
detect heliospheric structures via time-domain analysis of the
polarized intensity signal measured by three photometers on the spinning
spacecraft \citep{Jackson1986,Webb1987}.  Hardware on board
\emph{Helios} sorted detected photon events into accumulator
bins based on their temporal phase relative to the 1 Hz spacecraft spin. 
These bins were accumulated to yield sky brightnesses, including
polarization and color signals, on time scales of several hours.
These angularly separated photometric signals were used to
generate synoptic maps of the Thomson scattering surface brightness of the
solar wind \citep{Hick1991}, to estimate CME mass
\citep{JacksonWebb1995,Webb1995}, and even to
constrain coarse tomographic reconstructions of the
three dimensional structure of the heliosphere \citep{Jackson1995}.

In Section \ref{sec:Elementary}, below, we discuss
heliospheric imaging with polarized light in the same context as we
have previously with unpolarized light \citep[Paper I:][]{Howard2012},
including: basic theory; importance of the Thomson surface; a summary
description of instrument sensitivity, detectability, and signal-to-noise;
and discussion of two key scientific applications of polarized imaging.
Section \ref{sec:Forward-Modeling-of} moves to simulations of these
effects applied to non-infinitesimal features including CMEs and
corotating interaction regions (CIRs). In Section \ref{sec:Discussion}
we discuss the applied theoretical results and their implications
for future missions.

\section{Elementary Polarized Theory}\label{sec:Elementary}

Thomson scattering theory has been covered at great length by many
authors. Finding the polarized radiance\footnote{As in Paper I, we eschew the term ``brightness'' that has been used
confusingly in the literature to represent either radiance or intensity,
even though we continue using the \emph{B} and \emph{pB} abbreviations
(originally standing for ``brightness'' and ``polarized brightness'') for radiance.
} 
of a plasma illuminated by a
distributed object that subtends significant solid angle from the
scatter point (e.g. a solar coronal feature only a few solar radii
above the photosphere) requires geometrical integrals over the range
of solid angles of the incoming light, and is even more complicated in
the general case than is the same calculation for unpolarized light.
The problem has been treated in this general case by numerous authors
\citep[e.g.][]{vandehulst1950,Billings1966,Poland1976,DeForest1995,Howard2009}
as part of coronagraphic image intepretation. But in the heliospheric
case the Sun may be treated as a small object, greatly simplifying the
theory (Howard \&\ Tappin 2009; Paper I). Here, we re-derive the
polarized radiance formulae in this simpler case, and outline the consequences of its
form for different aspects of heliospheric imaging through polarizers.

\subsection{\label{sub:Polarized-Scattering-Basics}Polarized Scattering Basics}

%
\begin{figure*}[!tb*]
\begin{centering}
\includegraphics[width=5in]{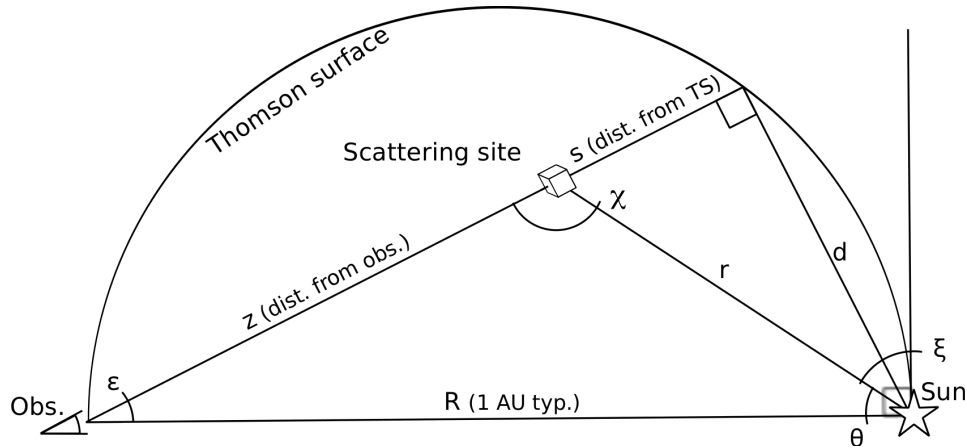}
\par\end{centering}

\caption{\label{fig:Observing-diagram-showing}Observing diagram shows relevant
angles for heliospheric imaging in the context of the Thomson scattering
geometry (from our Paper I). The line of sight with elongation $\varepsilon$
passes through the scattering site, making an angle of $\chi$ with
the radial from the Sun. Distance along the line of sight is denoted
$s$ when measured from the Thomson surface and $z$ when measured
from the observer.}

\end{figure*}
%

The elementary unpolarized theory of Thomson scattering begins with
the differential scattering cross-section for unpolarized light \citep{JacksonBook}:
\begin{equation}
\frac{d\sigma}{d\omega}=\sigma_{t}\left(1+\cos^{2}\chi\right),\label{eq:dsdw-unpolarized}
\end{equation}
where $d\sigma$ is the differential cross section for a single electron
to scatter unpolarized light into the solid angle $d\omega$, $\sigma_{t}\equiv r_{e}^{2}/2$
is half the square of the classical electron radius, and $\chi$ is
the angle of scatter, with $\chi=0$ for forward scatter. Thomson
scattering is absorption and re-radiation (in a dipole pattern) of
incident electromagnetic radiation. The electromagnetic waves can
be resolved into two polarized intensity or radiance components: $I_{\perp}$or
$B_{\perp}$ of light polarized perpendicular to the plane of scatter,
and $I_{\parallel}$ or $B_{\parallel}$ of light polarized in the
plane of scatter. In the circularly symmetric
solar case $B_{\perp}$ is also called the \emph{tangential} component and 
$B_{\parallel}$ is also called the \emph{radial} component.
In local observing coordinates with the Y axis aligned
radially outward from the Sun, the quantity $I_{\parallel}-I_{\perp}$
is the familiar ``Q'' Stokes parameter.  The two components are scattered
quite differently. The perpendicular component is scattered independently
of $\chi$, because of the circular symmetry of dipole radiation:
\begin{equation}
\frac{d\sigma_{\perp}}{d\omega}=2\sigma_{t},\label{eq:dsdw-perp}
\end{equation}
The factor of 2 reflects the fact that the perpendicular
polarization is only half of the original unpolarized beam -- while we
are now considering a  beam that is fully plane polarized perpendicular
to the plane of scatter. For the parallel component,
the electric field is simply projected with $\cos\chi$, so that the
overall intensity is  scaled by $\cos^{2}\chi$:
\begin{equation}
\frac{d\sigma_{\parallel}}{d\omega}=2\sigma_{t}cos^{2}\chi,\label{eq:dsdw-parallel}
\end{equation}
and it should be clear that Equation (\ref{eq:dsdw-unpolarized})
is simply the average of Equations (\ref{eq:dsdw-perp}) and (\ref{eq:dsdw-parallel}).
Coronagraphic and heliospheric imagery viewed through a radial/tangential
polarizer are thus quite different from the same imagery viewed with
no polarizing optics. It is most convenient to resolve observed
radiance features into an unpolarized radiance $B\equiv B_{\perp}+B_{\parallel}$
and an ``excess polarized radiance'' $pB\equiv B_{\perp}-B_{\parallel}$
\citep[e.g.][]{Fisher1981}. Because the propagators are linear, there
is a ``\emph{pB} scattering cross-section'' made from the difference
between the two polarized cross sections just as the unpolarized cross-section
is made from their sums:
\begin{equation}
\frac{d\sigma_{P}}{d\omega}=\sigma_{t}\left(1-\cos^{2}\chi\right)=\sigma_{t}\left(\sin^{2}\chi\right).\label{eq:dsdw-pb}
\end{equation}

\begin{figure*}[!tb]
\centering{}\includegraphics[width=5in]{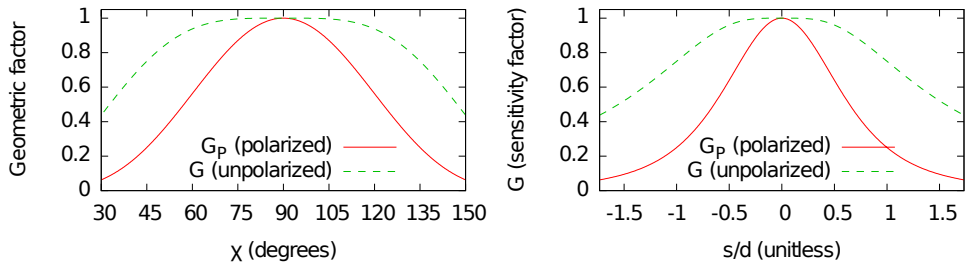}
\caption{\label{fig: G-vs-gp}Comparison of the \emph{pB} geometric factor
$G_{P}$ to the unpolarized \emph{B} geometric factor $G$ (derived
in Paper I) in both scattering angle $\chi$ and normalized distance
$s/d$ shows sharply peaked behavior for \emph{pB}. The sharp peak at $90\degr{}$
is due to the combined local maximum of illumination and the $\sin^{2}\chi$
dependence of the $pB$ scattering term; see Equations (\ref{eq:dsdw-pb})
and (\ref{eq:dB}).}
\end{figure*}

As in Paper I, we treat the Sun as a small object for the current 
  heliospheric case. This greatly simplifies the integral formulation developed by
  earlier authors.  The differential radiance is simply
  proportional to the mean solar radiance and the Sun's apparent size from
  the point of scatter:
\begin{equation}
d\left(pB\right)\equiv\frac{dP_{\perp}-dP_{\parallel}}{d\omega dA}=\sigma_{t}\left(\sin^{2}\chi\right)\left\{ \left(\frac{\pi r_{\odot}^{2}}{r^{2}}B_{\odot}\right)n_{e}\right\} ds,\label{eq:dB}
\end{equation}
 where $r_{\odot}$ and $B_{\odot}$ are the radius and mean radiance of the
 Sun, $ds$ is distance along a hypothetical line of sight, and $dA$ is
 surface area normal to that line of sight.  The only difference
 between the polarized radiance $pB$ and the radiance $B$ is that the
 $(1+\cos^{2}\chi)$ term from Equations~(1) and (2) of Paper I becomes
 the $\sin^{2}\chi$ term in our Equations (\ref{eq:dsdw-pb}) and
 (\ref{eq:dB}). Separating out the $\chi$-dependent portion of
 $d\left(pB\right)$ echoes Equation (6) of Paper I:
\begin{equation}
d\left(pB\right)=k_{TS}(\varepsilon)G_{P}(\chi)n_{e}(s,\varepsilon,\alpha)ds,\label{eq:Gp}
\end{equation}
 where $\varepsilon$ is elongation angle, $s$ is distance along
a line of sight, $\alpha$ is an azimuthal angle around the Sun in
the celestial sphere, and $n_{e}$ is the numerical electron density
across space. The in-plane geometrical values are summarized in Figure \ref{fig:Observing-diagram-showing}.
The functions $k_{TS}$ and $G_{P}$ serve to isolate the $\chi$
dependence of $pB$. Those last two terms are given by 
\begin{equation}
k_{TS}(\varepsilon)=\left(B_{\odot}\sigma_{t}\pi r_{\odot}^{2}\right)\left(R\,\sin\varepsilon\right)^{-2}\label{eq:kts}
\end{equation}
 and 
\begin{equation}
G_{P}=\sin^{4}\chi,\label{eq:G_P}
\end{equation}
where $k_{TS}$ includes the $\chi$-independent (and hence
$s$-independent) parts of $d\left(pB\right)$, the geometric factor
$G_{P}$ includes the $\chi$-dependent parts and again the geometric
quantities are as summarized in Figure
\ref{fig:Observing-diagram-showing}. $G_{P}$ includes both
illumination and scatter dependence on $\chi$, and is sharply peaked
around $\chi=90\degr{}$.  This is in contrast to the unpolarized
geometric factor $G$, discussed in Paper I, which has an
extraordinarily broad peak around $\chi=90\degr{}$, superosculating
the $f(\chi)=1$ line there. Figure \ref{fig: G-vs-gp} shows the
relationship between $G$ and $G_{P}$.  This relationship between the B
and pB signals' variance with angle has been known over 60 years
\citep{vanHouten1950} and has been used to interpret coronagraph
observations since at least the time of \emph{Skylab}
\citep[e.g.][]{Poland1976,Dere2005}. So far, it has neither been well
explored nor exploited in the wide-field (heliospheric) case, which is
considerably different from the coronagraph case (as described in
Paper I).

The two curves $G$ and $G_{P}$ are tangent at $\chi=90\degr{}$
because at $90\degr{}$ the Thomson scattered light is
fully polarized -- so $pB=B$ at that scattering angle. The second
derivative of $G_{P}$ is strongly negative at the peak, so that \emph{in
}pB\emph{ measurements there is no Thomson plateau}, and \emph{pB}
measurements are most sensitive to material close to the Thomson surface
(TS). In particular, \emph{pB} images are moderately well localized
to the vicinity of the TS.

Considering the scattering efficiency factors in $G$ and $G_{P}$
independently of illumination: the scattering efficiency for unpolarized
radiance $B$ is minimized on the Thomson surface, while the ``scattering''
effiency for excess polarized radiance $pB$ is maximized on the Thomson
surface.

%
%

\begin{figure*}
\begin{centering}
\includegraphics[width=6in]{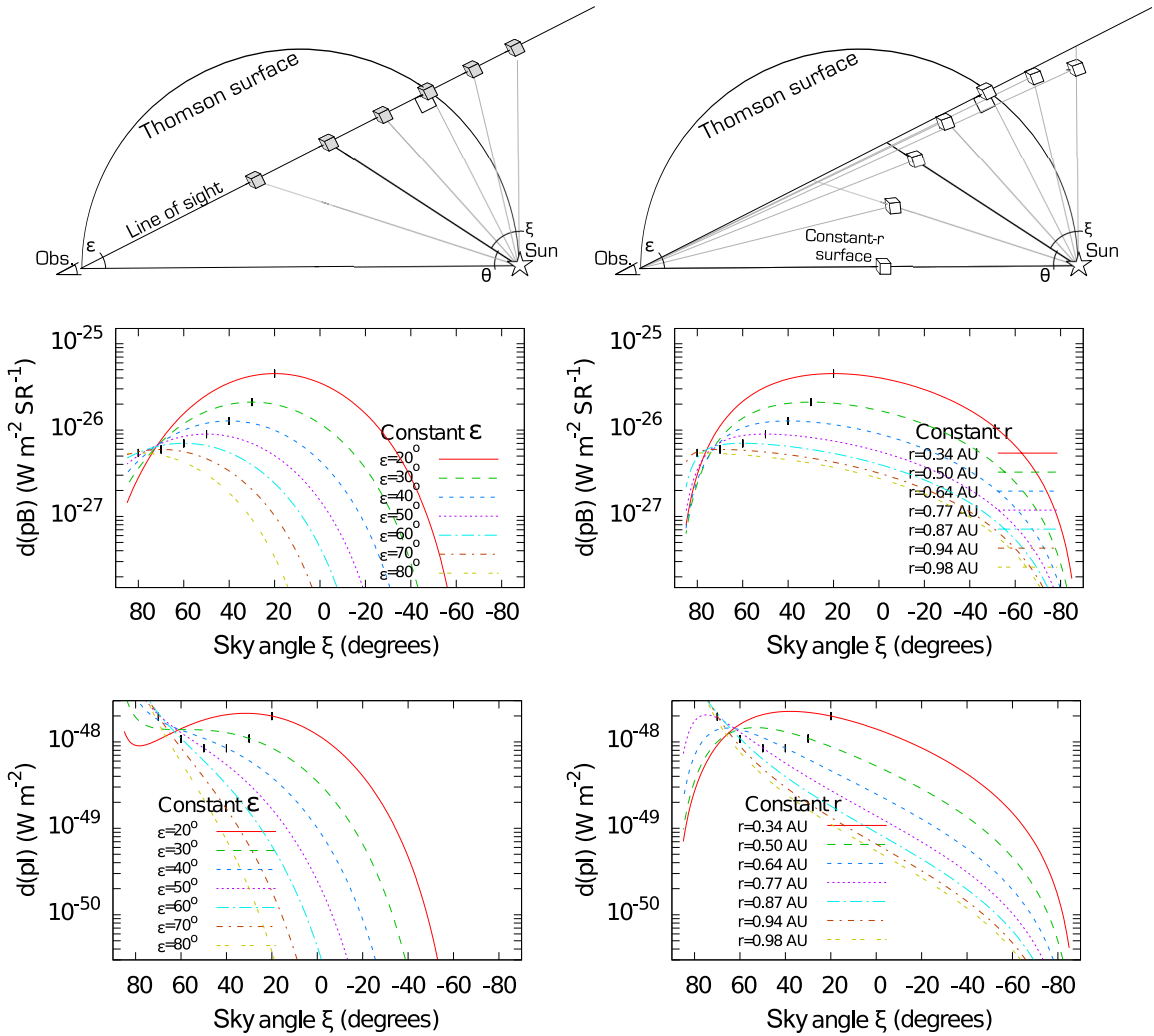}
\par\end{centering}

\caption{\label{fig:Thomson-scattering-effects}Thomson scattering effects
localize the \emph{pB} signal to the TS both at constant $\varepsilon$
and at constant $r$. At left: excess polarized surface brightness \emph{d(pB)}
and intensity \emph{d(pI)} at constant $\varepsilon$ show feature contrast variation
at a particular location in an instrument image plane.  At right: the same quantities
plotted at constant $r$ show how detectability varies with exit angle. In all 
plots, the intersection of each line with the TS is marked.  The TS marks a local
maximum in \emph{d(pB)} in both the constant-$\varepsilon$ and the constant-$r$ cases. 
At some (but not all) elongations there is a local maximum of \emph{d(pI)} near (but not on) the TS (lower
left).}

\end{figure*}

The polarized excess intensity (\emph{pI}) of a unit volume of material
is the polarized radiance integrated over its apparent size. Carrying
out the same operations as for Equation (9) of Paper I yields:
\begin{equation}
d\left(pI\right)\equiv pB\, d\Omega=\left\{ \frac{B_{\odot}\sigma_{t}\pi r_{\odot}^{2}}{R^{4}}\right\} \left[\frac{{\sin^{6}\chi}}{\sin^{2}\left(\varepsilon\right)\,\sin^{2}\left(\varepsilon+\chi\right)}\right]dN_{e}.\label{eq:pI}
\end{equation}
Both $d\left(pB\right)$ and $d\left(pI\right)$ for $dV=1m^{3}$
and $n_{e}=1m^{-3}$ are plotted at constant elongation $\varepsilon$
and at constant heliocentric distance $r$ in Figure \ref{fig:Thomson-scattering-effects},
versus the out-of-sky-plane angle (or ``sky angle'') $\xi$ (with
the understanding from Figure \ref{fig:Observing-diagram-showing}
that $\xi=\varepsilon+\chi-\pi/2$). Figure \ref{fig:Thomson-scattering-effects}
is directly comparable to Figure 5 of Paper I. The location of the TS is marked with
a single vertical bar across each line-of-sight plot. The $pB$-vs-$\xi$
curve (middle left) echoes the sharpness of the $G_{P}$ geometric
function in Figure \ref{fig: G-vs-gp}. Unlike the unpolarized radiance
$B$, there is no local minimum of $pB$ at the TS in the constant-$r$
case (upper right). 

Also unlike the unpolarized case, $pI$ has a local maximum for a given
size of feature for elongation angles smaller than 30\degr{} (green
dashed curve at lower left). At wide elongation angles, perspective
effects overwhelm the Thomson scattering and illumination efects, and
there is no local maximum. In all cases, perspective effects skew the
location of greatest polarized intensity away from the TS, although 
the greatest polarized radiance occurs on the TS. The 20\degr{} elongation
curve, at lower left, for example, shows a peak excess polarized
intensity about 10\degr{} closer to the observer than is the TS. At
constant $r$, there is a local maximum in excess polarized intensity
across $\chi$ at each possible exit angle.
This is in striking contrast to the unpolarized case explored in
  Paper I, in which we showed that there is no local maximum in
  intensity for features with small apparent size.


\subsection{\label{sub:Signal-to-Noise}Sensitivity and Signal-to-Noise Ratio}

As discussed in Paper I, detectability of features visible in \emph{$pB$}
depends not on their radiance but on their\emph{ }total intensity
(in this case, total polarized intensity) above a noise floor. Because
$pB$ and $pI$ are compound measurements (they are formed from the
difference between two radiances or intensities, respectively), the
intrinsic photon noise for $pI$ is greater than for a single measurement
of \emph{$I_{\perp}$} or $I_{\parallel}$. But in the heliospheric
case, intrinsic photonic noise in the intensity of the desired Thomson
scattered feature is small compared to the large background (and its
associated photon and other noise sources) that must be subtracted.
Even outside the brighter regions of the zodiacal light and the galaxy,
the surface brightness is dominated by the starfield \citep[e.g.][]{Jackson2010}.
Following the analysis in Paper I, we note that the background noise
against which features are measured is a random variable with approximately
constant distribution, with an approximately constant number of samples
per unit solid angle on the celestial sphere. This means that $N=L\Omega^{0.5}$,
where $N$ is the noise against which a feature is to be compared,
$\Omega$ is the solid angle subtended by it, and $L$ is an instrument-dependent
factor. It is thus possible to calculate \emph{a priori} how the signal-to-noise
ratio (SNR) varies with geometry for any small visual feature, up
to a normalization constant. 

\begin{figure*}[!tb]

\begin{centering}
\includegraphics[width=6in]{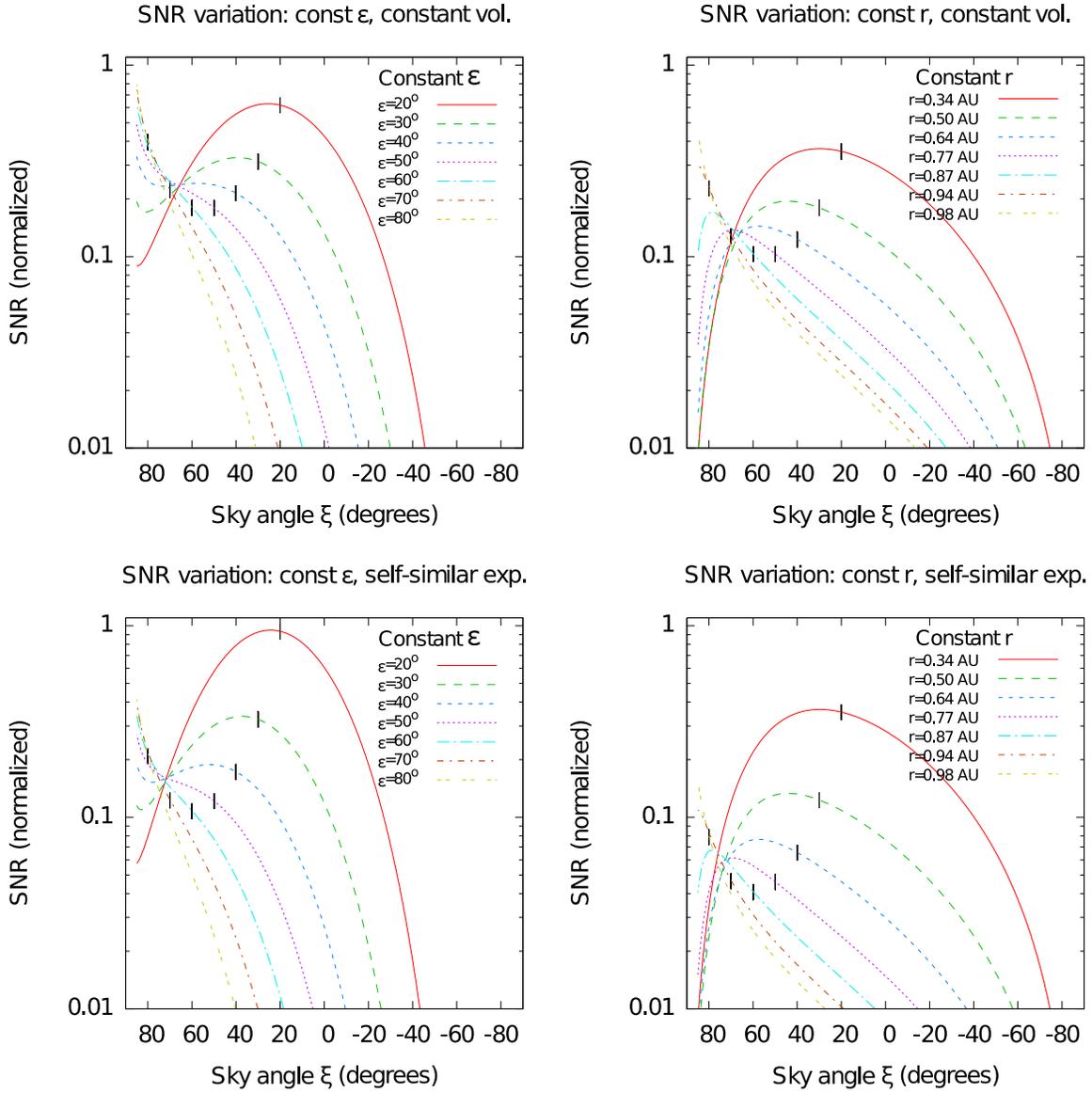}
\par\end{centering}

\caption{\label{fig:Signal-to-noise-ratio-variation}Signal-to-noise ratio
(SNR) variation for polarized intensity, vs.\  sky angle $\xi$ for
four cases: at left, constant-$\varepsilon$; at right, constant-$r$
At top: constant feature volume; at bottom: self-similar expansion.
In all cases, the SNR drops more rapidly than in the unpolarized case
described in Paper I. Observer-directed events have SNRs reduced by
a factor of order 10 compared to the ideal viewing angle. The curves
have been normalized with the same coefficients as Figure 6 of Paper
I, and are therefore directly comparable to their counterparts there.}
\end{figure*}

Figure \ref{fig:Signal-to-noise-ratio-variation} shows how the ratio
varies for different types of feature and different types of angular
comparison. The left two panels show line-of-sight comparison and the
right two panels show constant-radius comparison, as in Figure
\ref{fig:Thomson-scattering-effects}. The top two panels show behavior
of a feature with constant volume and mass as it propagates out from
the Sun, and the bottom two panels show the behavior of a
self-similarly expanding feature of constant mass as it propagates
outward, i.e. a feature that retains its shape but scales to larger
size with propagation as described by \citet{DeForest2012}.  Real
solar wind features typically fall between these two cases. The
self-similar expansion with constant mass is pessimistic far from the
Sun as most features expand laterally but not radially as they
propagate; and many dense features also accumulate mass through
snowplow effects \citep[e.g.][]{DeForest2012}. Similarly, the
constant-volume expansion is somewhat optimistic far from the Sun as
most features expand laterally to occupy approximately constant solid
angle relative to the Sun.

Because noise analysis is strongly dependent on specifics of the
instrument used to measure a signal, we confine ourselves (as in
Paper I) to describing only \emph{variation} of the SNR of a
hypothetical feature across location in the heliosphere,
\emph{ceteris paribus}.  Polarization measurements have additional
instrument-dependent noise sources that must be considered in
addition to the basics of aperture and integration time for an
unpolarized measurement.  In particular, an instrument that collects
individual polarized signals in sequence through a single polarizer
will incur noise from evolution of the signal and background between
the two exposures, while instruments that rapidly modulate or that
use dual-beam polarization do not, but may incur other sources of
noise.  These noise sources are dependent on the specific technology
and instrument used to detect polarization, and generally increase
the denominator of the SNR; but they do not affect the form of the variation
of SNR with feature-Sun distance.

Because of the instrument-dependence of the absolute SNR, we
normalized the curves in Figure
\ref{fig:Signal-to-noise-ratio-variation} to be near unity, to display
how the ratio varies for a particular class of feature. The
normalization constants for Figure
\ref{fig:Signal-to-noise-ratio-variation} are numerically the same as
the normalization constants we used in the analogous figure for
unpolarized signal (Figure 6 of Paper I), and the figures may
therefore be compared directly although the maximum value in Figure
\ref{fig:Signal-to-noise-ratio-variation} is not always
unity. Polarized SNR is dominated by the interplay of the inefficiency
of the \emph{pB }scattering far from the Thomson surface, and the
proximity effect in intensity as the feature approaches the observer.

The right-hand plots show, unsurprisingly, that Earth-directed features
are not clearly visible from Earth in $pB$ alone. Features close
to Earth are more readily detectable but that is a perspective effect:
the peaks in SNR at 60\degr{}-80\degr{} elongation in
all the plots in Figure \ref{fig:Signal-to-noise-ratio-variation}
are due to large apparent-size effects in infinitesimal features;
the SNR of real features will thus roll off when the feature-observer
distance $z$ shrinks to the same order as the size of the feature
itself: at smaller values of $z$, the small-apparent-feature approximation
$\Omega\propto z^{-2}$ no longer holds.


\subsection{\label{sub:Comparison-with-Unpolarized}Comparison with Unpolarized
Scattering: 3-D effects}

\begin{figure}[!tb]
\begin{centering}
\includegraphics[width=3in]{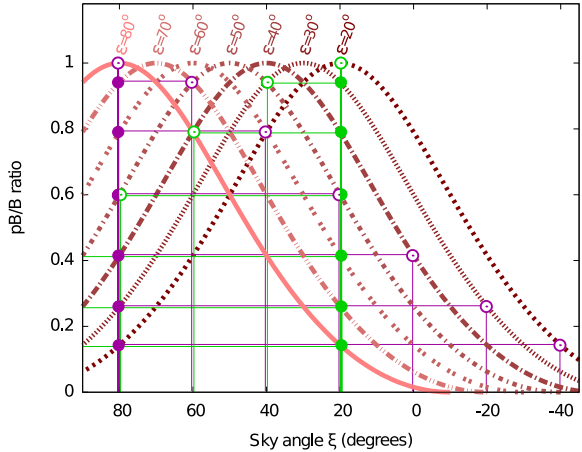}
\par\end{centering}

\caption{\label{fig:pB/B-ratio}The ratio of $pB$ to $B$ in small features
varies with sky angle $\xi$, and is plotted for several elongation
angles (dashed/dotted curves). Overplotted: two example features exit
the Sun at $\xi=20$\degr{} (green; nearly in the plane
of the sky) and $\xi=80$\degr{} (purple; just missing the observer),
passing through elongation angles between 20\degr{} and 90\degr{}.
At each elongation, the feature's $pB/B$ ratio corresponds to either
the real $\xi$ (bold vertical lines with solid circles) or to a ghost
location $\xi'$ that varies with elongation (horizontally extended
lines with open circles), which is dynamically implausible. See Figure
\ref{fig:Two-feature-trajectories} for a sketch.}
\end{figure}

\begin{figure*}[!tb]
\begin{centering}
\includegraphics[width=4.5in]{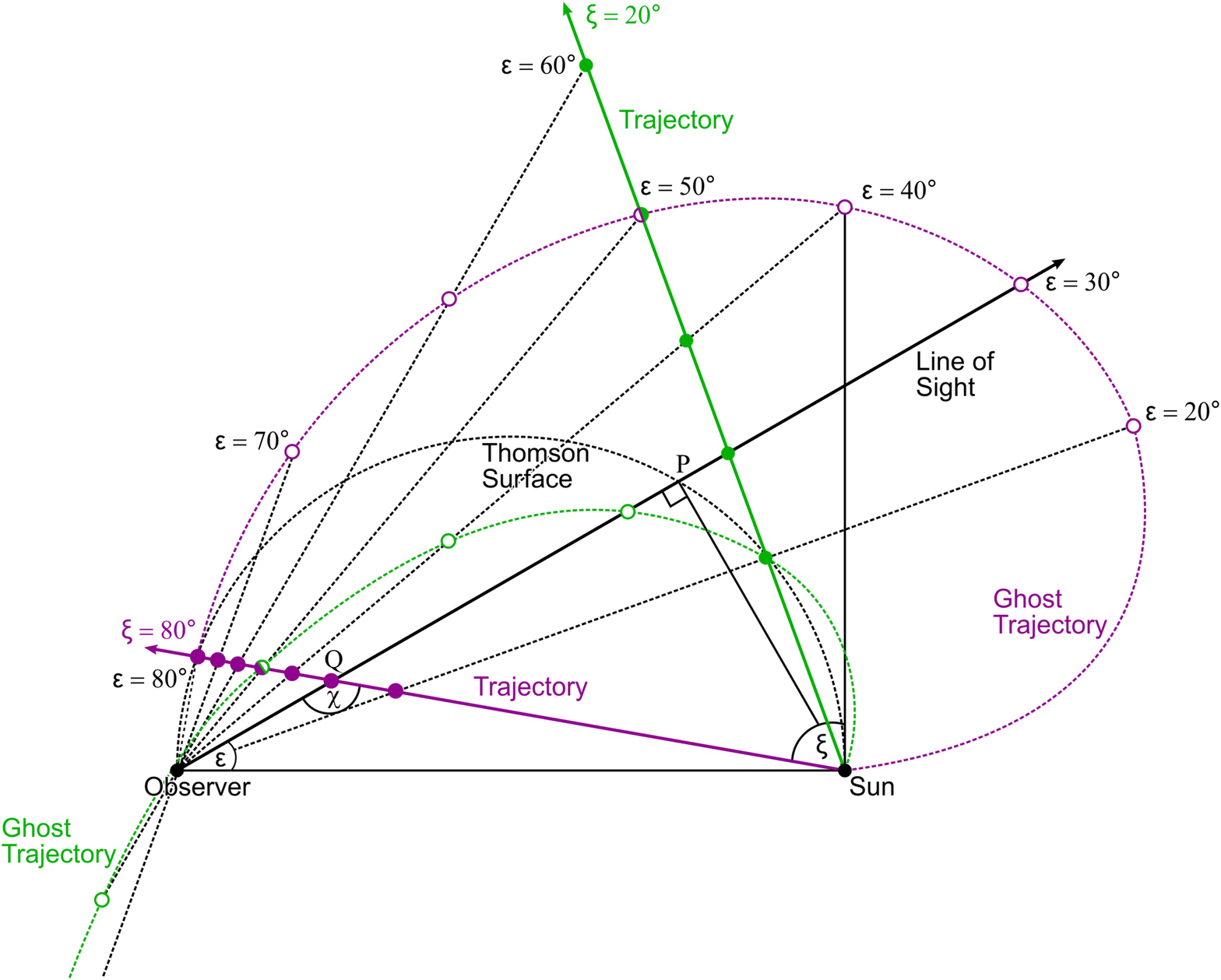}
\par\end{centering}

\caption{\label{fig:Two-feature-trajectories}Two feature trajectories show
how feature $\xi$ measurements relate to elongation angle $\varepsilon$.
The green ($\xi=20$\degr{}) and purple ($\xi=80$\degr{})
features propagate most of 1 AU on opposite sides of the Thomson surface.
The ratio $pB/B$ (or, equivalently, $pI/I$) in the feature is consistent
with two values of $\xi$ at each point - but the ``ghost trajectory''
 is unphysical. See Figure \ref{fig:pB/B-ratio}
for a plot of $pB/B$ for features along each line of sight in this
figure.}
\end{figure*}

Combining polarized and unpolarized imagery, with sufficient SNR,
enables location of features in 3-D with a single polarized
image pair. This technique has been explored extensively in
coronagraphs in recent years
\citep[e.g.][]{Moran2004,Dere2005,deKoning2011}.  Here we develop the
theory in the slightly different case of wide-field heliospheric imaging.  The ratio $pB/B$
(and its feature-averaged equivalent, $pI/I$ for a whole feature) is
just the ratio of the corresponding geometric factors:
\begin{equation}
\frac{pB}{B}=\frac{G_{P}}{G}=\frac{\sin^{2}\chi}{1+\cos^{2}\chi},\label{eq:pB/B}
\end{equation}
which has solutions: 
\begin{equation}
\chi=\textrm{acos}\left(\pm\sqrt{\frac{1-pB/B}{1+pB/B}}\right)\label{eq:3d-sol}
\end{equation}
or, giving the sky angle $\xi$ in terms of observed elongation angle
$\varepsilon$ and $pB/B$:
\begin{equation}
\xi=\varepsilon+\textrm{asin}\left(\pm\sqrt{\frac{1-pB/B}{1+pB/B}}\right).\label{eq:3d-sol-2}
\end{equation}
The relationship between $\xi$ and $pB/B$ is plotted in Figure \ref{fig:pB/B-ratio}
for two different features exiting the Sun at $\xi=20$\degr{}
(near the plane of the sky; green in the figure) and $\xi$=80\degr{}
(nearly directly at the observer; purple in the figure). The two branches
are on opposite sides of the Thomson surface, equally displaced from
it along the line of sight. In the coronagraphic case, the two branches
represent a permanent ambiguity, because $\varepsilon$ is nearly
zero and therefore the TS is in the plane of the sky. In the heliospheric
case, the curvature of the TS breaks the front/back asymmetry. While
any one heliospheric measurement of $pB/B$ cannot identify which
branch is occupied by a given feature, time series of observations
can: on the occupied branch, the trajectory is approximately inertial
and radial from the Sun, while on the opposite branch the inferred
``ghost trajectory'' includes large lateral accelerations that are 
reflected in the curved path in Figure \ref{fig:Two-feature-trajectories}.


Location of features in three dimensions requires propagating noise
in the photometry through the $\left(pI/I\right)$-to-$\xi$ conversion
process. Neglecting instrumental noise sources, the noise level of
the $\left(pI/I\right)$ measurement may be calculated by simple quadrature
combination of the $pI$ and \emph{I} noise sources. Treating the
noise in first order, the feature as compact and uniform-radiance,
and the two measurements (and noise samples) as independent, gives:
\begin{equation}
\Delta\left(\frac{pI}{I}\right)\approx\left(\frac{\Delta pI}{pI}\oplus\frac{\Delta I}{I}\right)\left(\frac{pI}{I}\right),\label{eq:3Dnoise}
\end{equation}
where $\oplus$ denotes addition in quadrature. The resulting noise
term can then be propagated through Equation (\ref{eq:3d-sol-2})
to identify error in $\xi$ determination for any given noise level
in the images. The two terms added in quadrature are the SNRs of the
$pI$ and $I$ measurements, respectively. This treatment glosses
over the fact that $pI$ is a compound measurement and hence incurs
two samples of the background noise, but that makes no difference
because: (a) the photon noise in the signal itself is considered to
be negligible compared to that of the background; and (b) if $pI$
and $I$ are assembled from the sum and difference of two intensities
obtained through polarizers (or through one rotating polarizer), there
are still only two samples of the background photon noise to be considered.

\begin{figure}
\begin{centering}
\includegraphics[width=3in]{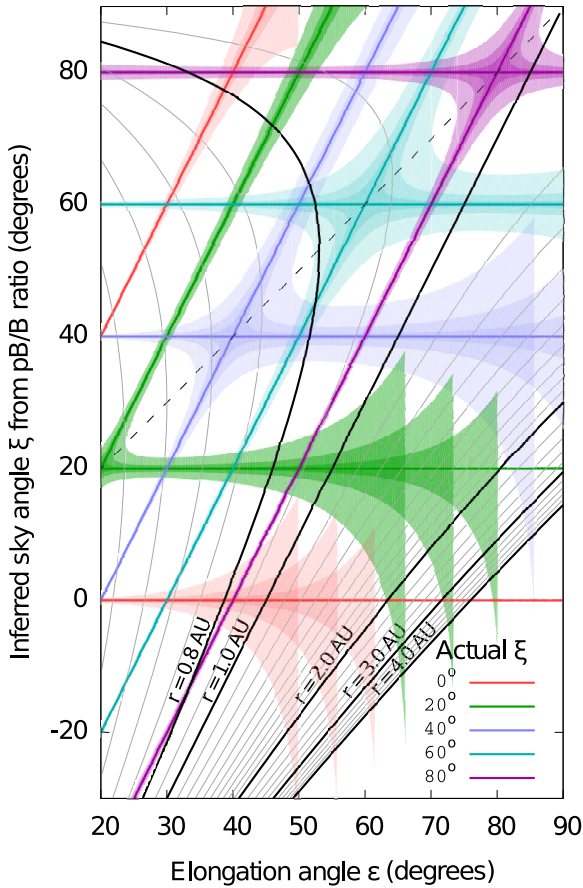}
\par\end{centering}

\caption{\label{fig:Inferred-sky-angle}Inferred sky angle $\xi$ vs.\  elongation
$\varepsilon$ for five small features with moderate SNR, using direct
measurement of the $pI/I$ intensity ratio. The inferred physical
(horizontal) and ghost (diagonal) trajectories are solid lines colored
by feature. Shapes around each line are $\xi$ error bars for self-similarly
expanding features with three SNRs at 0.8 AU: 10 (wide), 30 (medium),
and 100 (narrow). The green and purple features are also shown in
Figures \ref{fig:pB/B-ratio} and \ref{fig:Two-feature-trajectories}.
The Thomson surface ($\xi=\varepsilon$) is marked as a dashed black
line, and curves of constant radius from the Sun are marked as solid
black curves. }
\end{figure}

To understand the behavior of actual 3-D interpretation, we work from
the perspective of trying to determine the position of a particular
feature in radial motion, whose total intensity is measured with a
given SNR in the unpolarized channel. Figure \ref{fig:Inferred-sky-angle}
shows calculated position of an ensemble of five small features with
moderate SNR, using direct measurement of the $pI/I$ intensity ratio.

Two of the features in Figure \ref{fig:Inferred-sky-angle} correspond
to the two features followed by Figures \ref{fig:pB/B-ratio} and
\ref{fig:Two-feature-trajectories}. All of the features were treated
as self-similarly expanding, with an average SNR across each of
  the two polarized channels  of 10, 30, and 100 at 0.8 AU
from the Sun.  The three SNRs yield  three sets of
error bars on the inferred value of $\xi$; these error bars are
represented with the nested semitransparent curves around each $\xi$
trace. To find the SNR at each value of $r$, we used the same formulae
as for Figure \ref{fig:Signal-to-noise-ratio-variation} (and for the
$I$ measurement using the formulae for Figure 6 of Paper I). For
example, following the green trace we observe that a feature leaving
the Sun at $\xi=20\degr{}$ that happens to have a SNR of 10 at
elongation $\varepsilon=45\degr{}$ can be determined from a single
$pI/I$ measurement to have $\xi$ either between 15\degr{} and
25\degr{}, or between $67\degr{}$ and 73\degr{}. The same feature with
an SNR of 30 or 100 has similar inferred locations but tighter error
bars.

The shapes of the measurement loci in Figure \ref{fig:Inferred-sky-angle}
are dominated by the two branches of the $\xi$ calculation in Equation
(\ref{eq:3d-sol-2}). Since the features propagate radially, the correct
branch yields $\xi$ determinations that are constant as the feature
propagates outward through different elongations, while the ghost
trajectory (Figure \ref{fig:Two-feature-trajectories}) yields rapidly
changing $\xi$ angles versus elongation $\varepsilon$ as the feature
popagates. The intersection of the two branches occurs at the TS,
which is marked with a dashed black line. The plot thickens there,
primarily because the $pB/B$ ratio is independent of $\xi$ at the
TS. The two branches are symmetric about the Thomson surface because
of the symmetry of the $G$ and $G_{P}$ functions around $\chi=90\degr{}$
(Figure \ref{fig: G-vs-gp}).

All of the features in Figure \ref{fig:Inferred-sky-angle} show greatly
increased uncertainty toward the right of the plot, reflecting the
fact that the feature grows both fainter and farther from the observer
as it propagates away from the Sun. The $\xi$=0 curve, in particular,
has error bars that grow to over $\text{\ensuremath{\pm}45\degr}$
around the true measurement at around $\varepsilon=$60\degr{}
, as the SNR drops below unity: in that geometry, the feature is 2
AU from the Sun and 2.2 AU from the observer.

The feature in Figure \ref{fig:Inferred-sky-angle} with the least
uncertainty is at $\xi=80\degr{}$, headed nearly directly
toward the observer. For that geometry, perspective effects oppose
the illumination effects (though do not exactly cancel them) through
the mid portion of the trajectory. Hence, the SNR remains high throughout
the $\xi=80\degr{}$ trajectory, and the error bars remain
small except for the intrinsic uncertainty near the TS. 

The error bar shape is dominated by the computed SNR of the feature:
brighter, more compact features yield lower error bars, and fainter,
more diffuse features yield larger ones. In particular, we have used
self-similar expansion with constant $N_{e}$ (total electron count) in
the feature to scale the error bars as the features propagate outward;
this approach may be slightly pessimistic on the right-hand side of
the plot when compared to actual feature behavior, as discussed in
Section \ref{sub:Signal-to-Noise}.  

Real polarizing instruments may be subject to additional noise sources
that are more complex than the simple model we have used.  Calibration
errors between the two channels would introduce noise that scales with
the background and behaves functionally the same as the the noise
calculations we have used, and therefore do not affect Figure
\ref{fig:Inferred-sky-angle}.  Simultaneity errors could arise in
instruments that measure the two signals at slightly different times.
These types of error yield noise that is directly proportional to the
signal being measured and inversely proportional to the time scale of
its evolution, and therefore they would soften the curves in Figure
\ref{fig:Inferred-sky-angle}, by fading with distance from the Sun
even as the main noise source (background) increases with feature
size.

We conclude that $pB/B$ measurement of individual features' excess
radiance over background could be used to infer absolute exit angle
from the Sun with precision of a few degrees of angle, with SNR values
that are achievable with current instruments in $B$ alone 
\citep[as in][]{DeForest2011}.  While any one image yields ambiguous results
with two branches, time series of data reveal which of the two
branches is the correct one.  Near the TS, such measurements are very
imprecise because the slope of $pB/B$ with respect to the sky angle
$\xi$ is near zero - but radially propagating features necessarily
spend time both near and far from the Thomson surface. Those events of
primary interest to space weather prediction -- those with $\xi$ close
to $90\degr{}$ -- yield the most precise measurements of $\xi$, making
$pB/B$ measurements particularly interesting for this application. 

This analytic result applies only to \emph{small} features - i.e.
those for which the feature's characteristic length is negligible
compared to the distances to the Sun and observer, and which may
therefore be treated as point sources.  We discuss measurements of
large scale features briefly in Section \ref{sec:Forward-Modeling-of}
of this paper and at greater length (and more rigor) in Paper III of
this series \citep{Howard2012b}.


\subsection{Polarization effects and absolute background}

In coronagraphs, $pB$ measurements are used to reduce unwanted background
light from sources outside the coronagraph. In particular, the terrestrial
sky is quite bright near the Sun, and $pB$ measurements help to reduce
that contamination by subtracting two measurements with equal amounts
of background light \citep[e.g.][]{Lyot1933}. In terrestrial coronagraphs,
the technique is limited by polarization of the sky light at angles
surprisingly close to the Sun \citep{Leroy1972}. In space based coronagraphs,
the main persistent background is a combination of stray light from
within the instrument itself and sunlight scattered from microscopic
dust grains (the $F$ corona and zodiacal light; we use these terms
interchangeably). In the near-Sun field out to $\mbox{\ensuremath{\sim}}10\, R_{\odot}$
in the sky, the $F$ corona is nearly unpolarized, so $pB$ measurements
of the corona can provide an absolutely calibrated Thomson radiance
of features in the corona. Unpolarized coronagraphic measurements such 
as the primary synoptic data sets from \emph{SOHO}/LASCO
\citep{Brueckner1995}, and wide-field unpolarized heliospheric images
such as from \emph{STEREO}/HI \citep{Eyles2009}, cannot in principle separate
the background signal from a steady component of the Thomson radiance
under study, and therefore only report a ``feature excess'' radiance
\citep[e.g.][]{DeForest2012}. Here we discuss the feasibility of
similar absolute calibration of heliospheric Thomson scattered radiance
from a polarizing heliospheric imager.

In a spaceborne heliospheric imager, the background is a combination
of stray light from within the instrument itself, the unpolarized
starfield, and the $F$ corona, which ranges from $10^{-14}B_{\odot}$
to $10^{-13}B_{\odot}$ at elongations of interest to inner
heliospheric solar wind studies in the plane of the ecliptic. The zodiacal light dominates the sky background by photon count, and is
polarized by up to 20\% with a pole-to-ecliptic radiance variation of
about $4\times$ at 90\degr{} elongation from the Sun
\citep[e.g.][]{Dumont1976,Leinert1981,Leinert1998}.  In other words,
there is a background $pB$ signal of order a few
$\times10^{-14}B_{\odot}$ at moderate elongations from the Sun,
compared to the Thomson-scattering radiance of a few
$\times10^{-16}B_{\odot}$ in typical solar wind features. Thus, even $pB$
measurements require modeling of the diffuse background, just as do
direct $B$ measurements: individual constructed measurements include a
broad diffuse background that must be subtracted from the data to
yield ``feature-excess'' radiance in individual features.

Even with the presence of persistent diffuse background, $pB$ affords
techniques for indirect or ongoing-calibration methods to improve
the modeled background, which could permit detection of absolute Thomson
radiance based on the improved modeling. In particular, the \emph{Helios
}probes found that the zodiacal light is extraordinarily stable across
solar wind condition \citep{Richter1982a,Richter1982b} and even across
the whole solar cycle. \citet{Leinert1989} conclude that the secular
variation of the zodiacal light radiance is at most of order 1\% (limited
by the longevity of the \emph{Helios }photocells) with similarly stable
polarization properties. This is useful because, even with as much
as 1\% long-term variation of unpolarized radiance in the zodiacal
light, the stable polarization properties can be used to reduce the
variation by another 1-2 orders of magnitude. Hence, once well characterized
with a particular instrument, a zodiacal light model may be used to
produce a stable zero-point for $pB$ measurements of Thomson scattering
at the few $\times10^{-17}B_{\odot}$ level for the life of that instrument. 

Other types of wide-field unpolarized background are present in low-Earth
orbit \citep{Jackson2010}, including high altitude orbital speed
ram airglow and high altitude aurora \citep{Mizuno2005}. Both effects
are thought to be unpolarized as the mechanism for each is direct
atomic line emission, though to our knowledge neither has been tested. 
The aurora, as measured by SMEI, is comparable to the zodiacal light
when observed but is not universally present.  The airglow is universal at
lower altitudes than that of SMEI, and 3-4 orders of magnitude fainter than the 
zodiacal light at the 570km altitude of the Hubble Space Telescope \citep{Brown2000}.

We conclude that polarization can remove and/or stabilize at least
some important background effects, affording long-term remote measurement
of absolute density in the solar wind once a sufficient baseline of
measurements is made. While the dominant steady diffuse light source
(the zodiacal light) is slightly polarized, it has been measured to
be remarkably constant in both intensity and degree of polarization;
and the dominant variable diffuse light sources in low-Earth orbit
(aurora and ram airglow) are thought to be unpolarized and 
hence not to affect absolute pB measurements, but have not yet been characterized
to the precision required for such measurements.  Given recent successes
with separation of the heliospheric Thomson signal from F corona and stellar background
based entirely on image morphology \citep{DeForest2011} and the difference in 
image morphology between airglow and the K corona, it is plausible
that such measurements could be made even if these terms prove to be 
variably polarized.

In addition to these non-Thomson-scattering background signals,
there is a Thomson-scattering background to any particular feature
under observation, because the heliosphere and corona are optically
thin.  Just as in the corona, heliospheric CMEs and other individual
bright features can be many times brighter than the rest of the
Thomson-scattered heliospheric signal in unpolarized light
\citep{DeForest2011, DeForest2012}, and the Thomson background can
be neglected for these features; but 
Thomson-scattered background is important to analysis of faint
features, just as it is in coronagraph data.

\section{\label{sec:Forward-Modeling-of}Forward Modeling of Heliospheric
Imagery}

In order to gain a ready appreciation of the differences between how
solar wind disturbances appear in total power and polarized radiance,
we present here some simple simulations of solar wind features to
approximate the qualitative behavior of the polarization signal
from features whose extent is large.
The intent of these simulations is to reveal how
Section~\ref{sec:Elementary-Polarized-Theory} changes when the
features are not small - i.e. when integration, rather than simple
proportional scaling, is required. The $B$ and $pB$ signals from
large structures with nontrivial geometry contain contributions
from many different $\xi$ angles along the line of sight, and
therefore do not follow the simple inversion in Figure
\ref{fig:Inferred-sky-angle}. We use the simplest possible simulations
to reproduce the gross morphology of several typical heliospheric
structures to highlight how the overall feature shape influences
the polarization signal. 
Quantitative analysis of more realistically simulated
large scale features is presented in Paper III
\citep{Howard2012b}.

We used the generation codes of the Tappin-Howard model
\citep{Tappin2009} to compute synthetic sky maps for a number of
idealized disturbances.  The TH model is a code that extracts 3-D
parameters of CMEs by fitting simple geometric models to the evolution
of the CMEs as they propagate.  Briefly, a model homogeneous solar wind
density is assumed and the Thomson scattering formulae are integrated
in a spherical grid. Disturbances are added by multiplying the
relevant cells in the grid by an enhancement factor.  These can be
bent shells that simulate CMEs \citep{Howard2010}, or spiral
structures representing corotating interaction regions (CIRs).  For
the purposes of this study, to simplify comparison, we only consider
ecliptic plane cuts through those maps.

\subsection{Small-Scale (Blob or Puff)}

\label{sec:small-cme}
\begin{figure}
\centering \includegraphics[height=0.75\textheight]{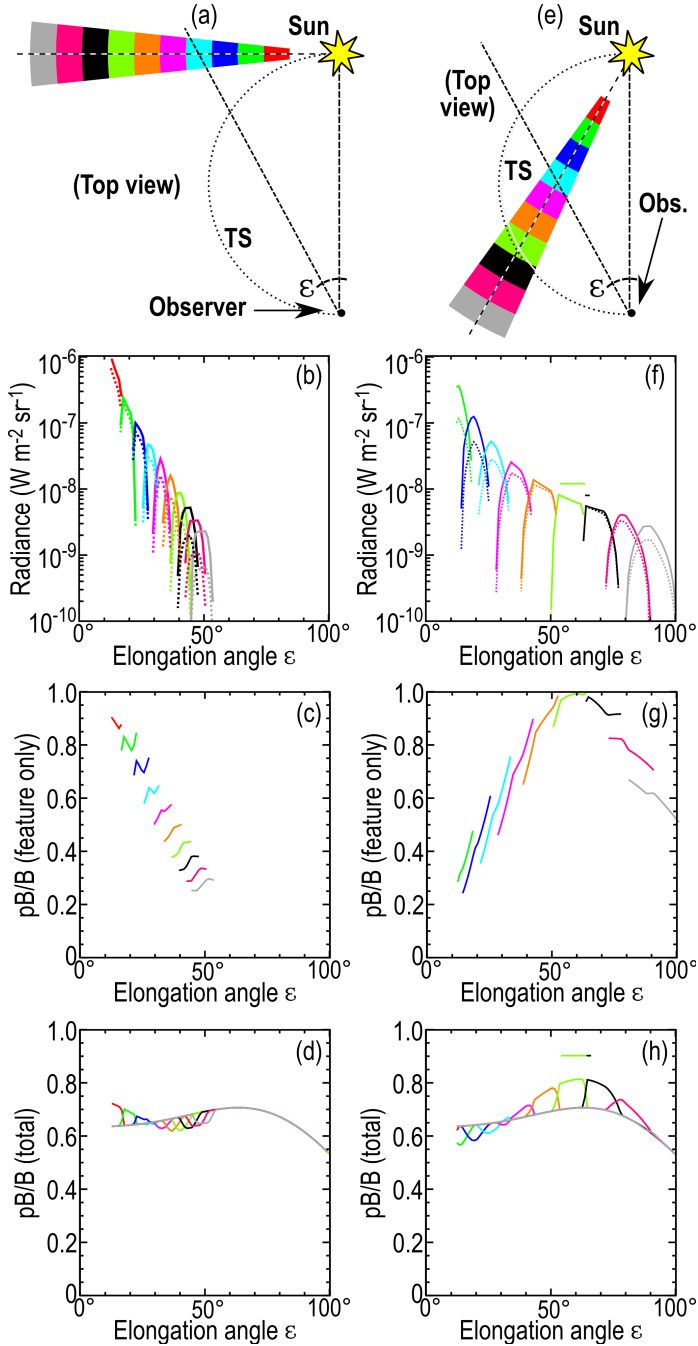} 
\caption{Radiance profiles through a small simulated transient. The
  left hand column (panels a-d) shows the case of the transient in the
  plane of the sky ($\xi=0\degr{}$), while the right column (panels
  e-h) shows the case where $\xi=60\degr{}$. The top row (a and e)
  shows the locations at which the transient radiance was computed:
  each color corresponds to a different radius; the dashed lines show
  the central axis of the transient and the Sun-Observer line.  The
  second row (b and f) shows the radiance excess as a function of
  elongation for each radius: the total power is solid; polarized
  power is dashed; the overbars indicate the elongation ranges within
  which the line of sight cuts the Thomson surface inside the
  transient.  The third row (c and g) shows the ratio of feature
  excess \emph{pB} to feature excess \emph{B} as a function of
  elongation. The fourth and final row (d and h) shows the ratios of
  total $pB$ to $B$ at the feature location (without background
  subtraction).}

\label{fig:small-scale} 
\end{figure}

The first disturbance considered was a small arc-shaped transient with
a thickness of 0.1~AU, and a cone angle of 6\degr{} (12\degr{} total
extent) in latitude and longitude, and an enhancment factor of 5 above
the ambient solar wind density. The enhancement factor is typical of
large CMEs \citep[e.g.][]{Liu2010,Mostl2012,Lugaz2012}.
The radiance was computed every 0.1~AU from 0.3-1.2~AU. We ran the
full simulation for two longitudes in heliocentric observing
coordinates, with the center of the CME at L=90\degr{}
(i.e. $\xi=0\degr{})$ and L=30\degr{} ($\xi=60\degr{}).$ The locations
of the CME are shown in Figure~\ref{fig:small-scale} (panels a and e).
Each colored sector corresponds to a location at which we computed the
radiances. In panels b and f we show the excess radiance (i.e.\ the
radiance with the quiet-wind values subtracted) for both total power
and polarized radiance. In panels c and g we show the ratio of the
polarized to total radiance excess , and in panels d and h the ratio
of polarized to total radiance. In all of the plots, the colors of the
traces match those in the schematic in panels a and e. At elongations
where the line of sight intersects the TS inside the transient, an
overbar is placed on the plots.

This small transient can be considered to approximate a blob of plasma,
and thus provides us with a clear view of the differences between
observing in total power and observing in polarized radiance.

The most obvious characteristic of the excess radiance plots is the
fall-off of radiance as the transient moves further from the Sun.
Although the derivative of signal with respect to elongation
$\varepsilon$ is much steeper at $\xi=0$\degr{} (left column) than at
$\xi=60$\degr{} (right column), that difference is primarily
geometrical.  Comparing the two sky-angle cases at similar radii
rather than similar elongations yields a far smaller differences
between the two cases, as might be expected from Figure
\ref{fig:Thomson-scattering-effects}.

Turning to the polarization fraction of the radiance excesses (panels
c and g of Figure~\ref{fig:small-scale}), the $pB$ signal falls off
rapidly with elongation $\varepsilon$ for the limb transient
($\xi=0$\degr{}) at left.  But for the near-observer transient
($\xi=60$\degr{}) at right, polarization increases to a maximum near
$\varepsilon=60$\degr{}.  Both effects may be expected from Figure
\ref{fig:Thomson-scattering-effects}.  For both transient longitudes
considered here, the signal from the transient at its closest to the
Thomson surface has a degree of polarization about 4 times greater
than at its furthest. 

As might be expected from the differential analysis in Section
\ref{sub:Polarized-Scattering-Basics}, when the transient is far from
the TS and its polarization is thus less than that of the quiet solar
wind, the polarization of the total received light is reduced. When it
is close to the TS the polarization is increased, reaching 80\%\ for
the transient directed 60\degr{} from the plane of the sky as that
transient crosses the TS.

\subsection{Large-Scale (CME)}

\label{sec:large-cme}

\begin{figure}
\centering \includegraphics[height=0.75\textheight]{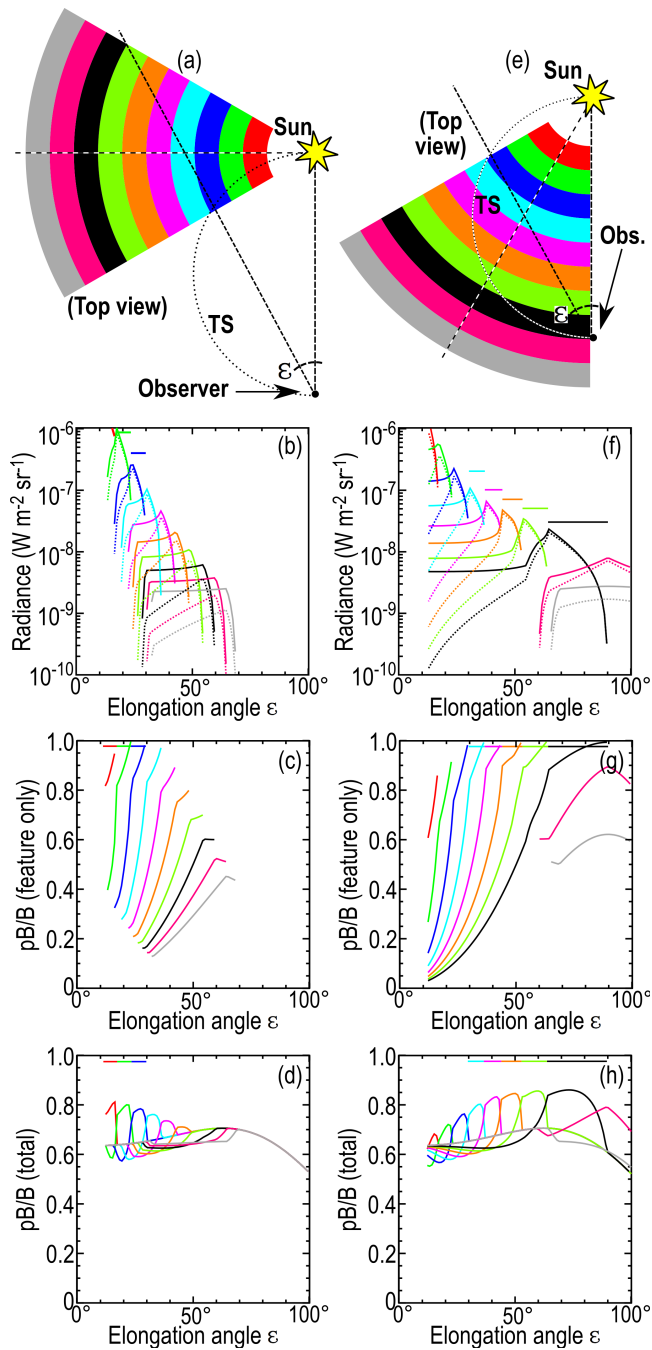} 
\caption{Display of the radiances from the large CME simulation. The format
is identical to Figure~\ref{fig:small-scale}.}

\label{fig:large-scale} 
\end{figure}

We repeated the simulations of Figure \ref{fig:small-scale} for a
  simplified conical CME with a longitudinal cone angle of 30\degr{},
and otherwise identical parameters. The results from this
cone are shown in Figure~\ref{fig:large-scale}, in the
same format as was used for Figure~\ref{fig:small-scale}.

As expected the CME appears much broader in
elongation.   The degree of polarization shows generally similar trends,
reaching a maximum near the Thomson surface. 
Where the CME first intersects the TS, the highest degree
of polarization (approaching 100\%) occurs at the leading edge of the
CME, thus producing a kind of intrinsic edge enhancement. The trailing
edge, which is remote from the TS, is de-emphasized,
in some cases by well over an order of magnitude. The CME directed at
30\degr{} always extends to the smallest elongations computed
(12\degr{}) when it lies between the Sun and the observer. This is
because this CME has a grazing impact with the observer. The extension
appears as the flat tails in the unpolarized radiance (solid) traces
in Figure~\ref{fig:large-scale}f.  In the polarized (dashed) traces,
on the other hand, the radiance falls off rapidly as the scattering
approaches pure forward scatter, which is unpolarized.

We conclude that the polarization signal from typical CMEs is not
completely spoiled by dilution due to the large angular extent of the
CMEs or the existence of a background solar wind. Further analysis is
required (and is presented in our Paper III of this series) to
determine how to extract CME position in three dimensions from the
polarization signal.

\subsection{Corotating interaction region.}

\label{sec:cir}

\begin{figure}
\centering \includegraphics[height=0.75\textheight]{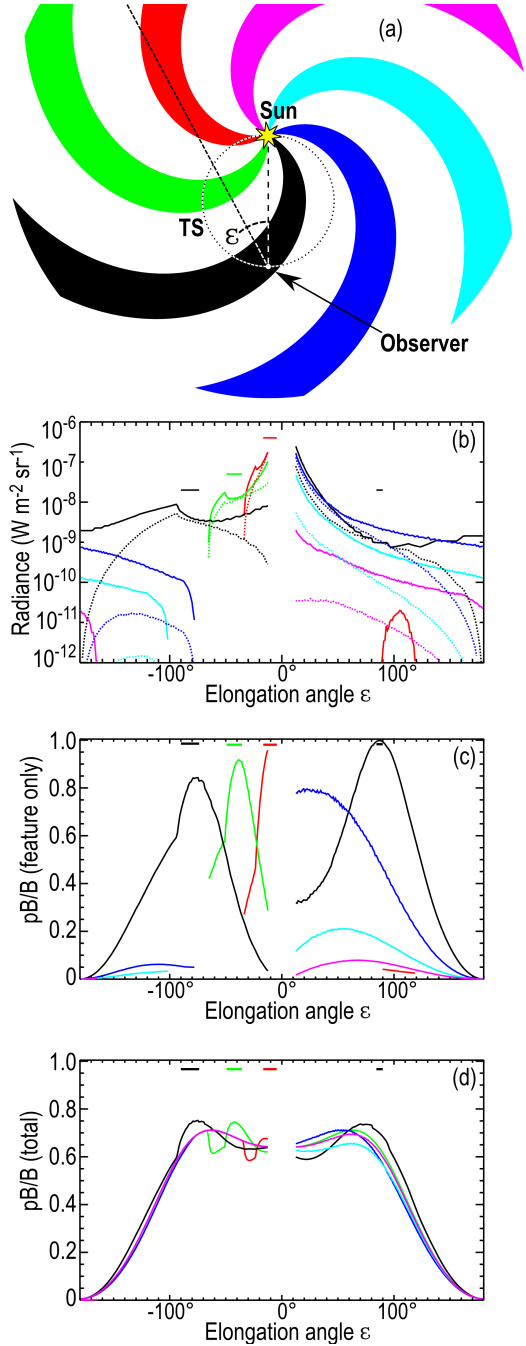} 
\caption{Display of the radiances from a simulated CIR. The format is similar
to that of Figure~\ref{fig:small-scale}, with the coloured spirals
representing the locations of the CIR at which the simulations were
performed. Since the CIR moves from East (left) to West (right) and
is not symmetrical, we show both sides of the Sun in the plots. Also
the computations are continued all the way to 180\degr{} elongation.
The elongation axis is reversed so that the traces appear in the same
orientation as on the sky. Note that computations are not performed
at elongations less than 12\degr{}.}

\label{fig:time-stat} 
\end{figure}

The third disturbance we simulated to generate intuition about
  the effects of geometry on pB was a spiral structure corresponding
to a 400$\mathrm{km\, s^{-1}}$ CIR in the plane of the
  ecliptic. We treated the CIR as a dense spiral
  structure with an extent in longitude of 24\degr{}, and increased
linearly in density to an enhancement factor of 3 at 1.2~AU, beyond
which it was of constant density. The results are summarized in
Figure~\ref{fig:time-stat} in a similar format to
Figure~\ref{fig:small-scale}, however here we show all elongations
from -180\degr{} to +180\degr{} (with negative ``elongations''
  corresponding to angle west of the Sun in the ecliptic, and positive ``elongations''
  corresponding to angle east of the Sun in the ecliptic). The computations were
carried out at longitudes of the leading edge of the base of 303, 3,
63, 123, 183 and 243\degr{} relative to the observer (the 3\degr{}
offset is a computational convenience). These correspond to
approximately -8.5, -4, +0.5, +5, +9.5 and +14 (-13) days from the
arrival of the CIR at the observer.

Again,  at elongations where the line
of sight intersects the TS inside the CIR or close to it, the polarization
is high (approaching 100\%), while at elongations where the line of sight
intersects the TS far from the CIR the polarization is far less pronounced even
if the unpolarized signal is strong.

A major difference in polarization behavior between the CIR and
  the earlier two cases is that when the CIR is growing in the
  eastern part of the image (i.e. approaching in the
east), the polarization at the leading edge is
persistently low (Figure~\ref{fig:time-stat}c and \ref{fig:time-stat}d), while
  the CME and compact structure display a maximum in polarization if
  and when they pass through the TS. This is a direct result of the
geometry of a CIR:  the tangential contact between the line of
sight and the CIR is dependent on the spiral angle of the CIR, and
  is always far from the TS; whereas
for most CMEs, when inside 1~AU, the tangent point is generally close
to the TS. See Figure~10 of Paper I for plots showing the
distance of the CIR leading edge from the TS.

We conclude that, although CIRs should be visible in polarized as
  well as unpolarized light, they behave differently than other types
  of feature because of their stationary, spiral shape.  In
  particular, the polarization behavior of the front of a CIR is quite
  different from that of a localized structure such as a CME.  CIRs
  should thus prove directly recognizable by inspection of polarized
  imges, even without detailed inversion for three-dimensional
  structure.

\section{\label{sec:Discussion}Discussion}

We have developed an appropriately-simplified theory of polarized
Thomson-scattered imaging in the heliosphere, and shown that sensitivity
of $pB$ measurements to electron density is well localized along
the line of sight at the TS, in contrast to the Thomson plateau observed
via unpolarized light ($B$). For unpolarized detection, the illumination
function and scattering efficiency have equal and opposite second
derivatives at the TS, leading to the Thomson plateau; in the case
of excess polarized radiance detection they have equal second derivatives
at the TS, leading to a sharper maximum in intensity than would be
observed via an $s-$scattering process with no angular dependence. 

The difference in spatial kernel between $pB$ and $B$ enables location
of individual solar wind features in three dimensions, and we have
developed a theory of small feature location including the importance
of SNR and kinematic effects in determining the precision to which
location may be measured. The curvature of the TS breaks the front/back
asymmetry that hinders attempts to accomplish the same thing in coronagraph
images near the Sun. Features far from the TS are more readily located
in three dimensions than are features near the TS, because the $pB/B$
ratio varies more with the sky angle $\xi$ far from the TS. An instrument
that could measure the two relevant polarizations with comparable
SNR to that of the \emph{STEREO}/SECCHI imagers could locate the exit
angle of small features within well under 10\degr{}, greatly
enhancing interpretation of the remote solar wind signal and potentially
improving space weather prediction through direct location of Earth-directed
features that are hard to measure geometrically \citep[e.g.][]{Lugaz2010}. 

The $pB$ signal is a differential signal that, by construction,
eliminates unpolarized bright features from the field of view. This
effect has been used historically in coronagraphs to remove stray
light and sky light from the Thomson scattering signal. Direct
application in the heliosphere is more complex because the zodiacal
light is polarized far from the Sun and hence is not removed by the
\emph{pB} calculation in the wide-field heliospheric imaging
case. However, the zodiacal light is extremely stable both in degree
of polarization and in overall intensity, leading to the possibility
of direct absolute measurements of the Thomson scattering signal using
long baselines and \emph{in situ} measurement of the average wind
density; the additional stability of the
polarization ratio further reduces the
residual background variation below what may be achieved with a single
measurement. Other diffuse light sources from near Earth include
orbital ram airglow and high altitude aurora, both of which we
anticipate to be unpolarized and hence invisible in \emph{pB.}

We have performed simple forward modeling of geometric structures
similar to known solar wind structures to develop intuition about how
a polarized heliospheric imager, if developed, will respond to those
features. We find that one may expect significant, unambiguous $pB$
signatures of the 3-D location of large features, although analysis of
such features is more complex than compact features whose geometry may
be neglected. Techniques for, and the limits
  of, large scale feature location in 3-D from $pB$ measurements
  require more detailed treatment and will be covered in the third
paper of this series.

\acknowledgements{ }

Support for this work is provided in part by the NSF/SHINE Competition,
Award 0849916 and the NASA Heliophysics program through grant NNX10AC05G.
SJT is supported at NSO by the USAF under a Memorandum of Agreement.
The authors gratefully acknowledge helpful discussions with L. DeForest,
C. Leinert, D. McComas, and A. Steffl. Much of the numerical analysis
for this work was performed with the community-developed Perl Data
Language (http://pdl.perl.org).


\begin{thebibliography}{DeForest, Howard \&{} McComas(2012)}
\bibitem[Billings(1966)]{Billings1966} Billings, D.~E., 1966, A
Guide to the Solar Corona, Academic Press, San Diego.

\bibitem[Brown et~al.(2000)]{Brown2000}
Brown, T.~M., Kimble, R.~A., Ferguson, H.~C., Gardner, J.~P., Collins, N.~R., and Hill, R.~S. 2000: Astron. J. 120, 1153.

\bibitem[Brueckner et~al.(1995)]{Brueckner1995}Brueckner, G.~E.,
Howard, R.~A., Koomen, M.~J., Korendyke, C.~M., Michels, D.~J.,
Moses, J.~D., Socker, D.~G., Dere, K.~P., Lamy, P.~L., Llebaria,
A., Bout, M.~V. Schwenn, R., Simnet, G.~M., Bedford, D.~K., \&
Eyles, C.~J. 1995: Sol. Phys. 162, 357.

\bibitem[Buffington et~al.(2009)]{Buffington2009} Buffington, A.,
Bisi, M.~M., Clover, J.~M., Hick, P.~P., Jackson, B.~V., Kuchar,
T.~A., \& Price, S.~D., 2009, Icarus, 203, 124.

\bibitem[Crifo et~al.(1983)]{Crifo1983} Crifo, F., Picat, J.~P.,
\& Cailloux, M., 1983, Solar Phys., 83, 143.

\bibitem[Davis et~al.(2009)]{Davis2009} Davis, C.~J., Davies, J.~A.,
Lockwood, M., Rouillard, A.~P., Eyles, C.~J., \&\ Harrison, R.~A.,
2009, Geophys. Res. Lett., 36, L08102, doi:10.1029/2009GL038021.

\bibitem[DeForest(1995)]{DeForest1995}DeForest, C.E. 1995, High Resolution
Multi-Spectral Observations of Solar Coronal Open Structures: Polar
and Equatorial Plumes and Rays, Ph.D. thesis, Stanford University.

\bibitem[DeForest et~al. (2011)]{DeForest2011}DeForest, C.E.,
Howard, T.A., and Tappin, S.J. 2011: Astrophys. J. 738, 103.

\bibitem[DeForest et~al. (2012)]{DeForest2012}DeForest,
C.E., Howard, T.A., and McComas, D.M. 2012: Astrophys. J. 745, 36.

\bibitem[DeMastus et~al.(1973)]{DeMastus1973} DeMastus, H.~L., Wagner,
W.~J., \&\ Robinson, R.~D., 1973, Solar Phys., 100, 449.

\bibitem[de Koning \&\ Pizzo(2011)]{deKoning2011} de Koning, C.~A.,
\& Pizzo, V.~J., 2011, Space Weather, 9, S03001, doi:10.1029/2010SW000595.

\bibitem[Dere et~al.(2005)]{Dere2005} Dere, K.~P., Wang, D., and Howard, R., 2005, Astrophys. J. L. 620, 119.

\bibitem[Dumont \&\ Sanchez(1976)]{Dumont1976}Dumont, R. and Sanchez,
F. 1976: Astron. Astrophys. 51, 393.

\bibitem[Eyles et~al.(2003)]{Eyles2003} Eyles, C.~J., Simnett, G.~M.,
Cooke, M.~P., Jackson, B.~V., Buffington, A., Hick, P.~P., Waltham,
N.~R., King, J.~M., Anderson, P.~A., \&\ Holladay, P.~E., 2003,
Solar Phys., 217, 319.

\bibitem[Eyles et~al.(2009)]{Eyles2009} Eyles, C.~J., Harrison,
R.~A., Davis, C.~J., Waltham, N.~R., Shaughnessy, B.~M., Mapson-Menard,
H.~C.~A., Bewsher, D., Crothers, S.~R., Davies, J.~A., Simnett,
G.~M., Howard, R.~A., Moses, J.~D., Newmark, J.~S., Socker, D.~G.,
Halain, J.-P., Defise, J.-M., Mazy, E., \&\ Rochus, P., 2009, Solar
Phys., 254, 387.

\bibitem[Fisher et~al.(1981)]{Fisher1981}Fisher, R.~R., Lee, R.~H.,
MacQuieen, R.~M., and Poland, A.~I., 1981, Applied Optics, 20, 1094.

\bibitem[Gosling et~al.(1975)]{Gosling1975} Gosling, J.~T., Hildner,
E., MacQueen, R.~M., Munro, R.~H., Poland, A.~I., \&\ Ross, C.~L.,
1975, Solar Phys., 40, 439.

\bibitem[Harrison et~al.(2008)]{Harrison2008} Harrison, R.~A., Davis,
C.~J., Eyles, C.~J., Bewsher, D., Crothers, S.~R., Davies, J.~A.,
Howard, R.~A., Moses, D.~J., Socker, D.~G., Newmark, J~S., Halain,
J.-P., Defise, J.-M., Mazy, E., Rochus, P., Webb, D.~F., \&\ Simnett,
G.~M., 2008, Solar Phys., 247, 171.

\bibitem[Hick et~al(1991)]{Hick1991} 
Hick, P., Jackson, B.~V., and Schwenn, R. 1991: 
Astron. Astrophys. 244, 242.

\bibitem[Hildner et~al.(1975)]{Hildner1975} Hildner, E., Gosling,
J.~T., MacQueen, R.~M., Munro, R.~H., Poland, A.~I., \&\ Ross,
C.~L., 1975, Solar Phys., 42, 163.

\bibitem[Howard \&\ DeForest(2012)]{Howard2012} Howard, T.~A., \&
DeForest, C.~E., 2012, Astrophys. J. 752, 130 (Paper I of this series).

\bibitem[Howard et~al.(2012)]{Howard2012b}Howard, T.~A., DeForest,
C.~E., Tappin, S.~J., \& Odstrcil, D. 2012, Paper III, submitted to Astrophysical Journal.

\bibitem[Howard, DeForest \&{} Reinard(2012)]{Howardetal2012} Howard,
T.~A., DeForest, C.~E., \&\ Reinard, A.~A., 2012, Astrophys. J.,
accepted.

\bibitem[Howard \&\ Tappin(2009)]{Howard2009} Howard, T.~A., \&
Tappin, S.~J., 2009, Space Sci. Rev., 147, 31.

\bibitem[Howard \&\ Tappin(2010)]{Howard2010} Howard, T.~A., \&
Tappin, S.~J., 2009, Space Weather, 8, S07004.

\bibitem[Howard et~al.(2006)]{Howard2006} Howard, T.~A., Webb, D.~F.,
Tappin, S.~J., Mizuno, D.~R., \&\ Johnston, J.~C., 2006: J. Geophys.
Res., 111, A04105, doi:10.1029/2005JA011349.

\bibitem[Jackson(1962)]{JacksonBook} Jackson, J.~D.,1962: ``Classical
Electrodynamics'', John Wiley \&\ Sons, New York.

\bibitem[Jackson(1986a)]{Jackson1986} Jackson, B.~V. 1986a: 
Adv. Sp. Res. 6, 307.

\bibitem[Jackson \& Froehling(1995)]{Jackson1995} 
Jackson, B.~V. and Froehling, H.~R. 1995: 
Astron. Astrophys. 299, 885.

\bibitem[Jackson \& Webb(1995)]{JacksonWebb1995}
Jackson, B.~V. and Webb, D.~F. 1995:
Proc. Sol. Wind 8, 97.

\bibitem[Jackson et~al.(2010)]{Jackson2010}Jackson, B.~V., Buffington,
A., Hick, P.~P., Bisi, M.~M., and Clover, J.~M. 2010: Solar Physics
265, 257.

\bibitem[Koomen et~al.(1975)]{Koomen1975} Koomen, M.~J., Detwiler,
C.~R., Brueckner, G.~E., Cooper, H.~W., \&\ Tousey, R. 1975: Appl.
Opt. 14, 743.

\bibitem[Leinert et~al.(1975)]{Leinert1975}
Leinert, C., Link, H., Pitz, E., Salm, N., and Knueppelberg, D. 1975: 
Raumfahrtforschung 19, 264.

\bibitem[Leinert et~al.(1981)]{Leinert1981}Leinert, C.,
Richter, I., Pitz, E., and Planck, E. 1981: Astron. Astrophys. 103,
177.

\bibitem[Leinert \&\ Pitz(1989)]{Leinert1989}Leinert, C., \& Pitz,
E. 1989: Astron. Astrophys. 210, 399.

\bibitem[Leinert et~al.(1998)]{Leinert1998}Leinert, C. et al. 1998:
Astron. Astrophys. Suppl. Ser. 127, 1.

\bibitem[Leroy et~al.(1972)]{Leroy1972}Leroy, J.~L.,
Muler, R., and Poulain, P. 1972: Astron. Astrophys. 17, 301.

\bibitem[Liu et~al.(2010)]{Liu2010}
Liu, Y., Davies, J.~A., Luhmann, J.~G., Vourlidas, A., Bale, S.~D., 
\&\ Lin, R.~P., 2010, Astrophys. J. Lett., 710, L82.

\bibitem[Lugaz(2010)]{Lugaz2010} Lugaz, N. 2010: Solar Phys. 267,
411.

\bibitem[Lugaz et~al.(2010)]{Lugaz2009} Lugaz, N., Hernandez-Charpak,
J.~N., Roussev, I.~I., Davis, C.~J., Vourlidas, A., \&\ Davis,
J.~A. 2009: Astrophys. J. 715, 499.

\bibitem[Lugaz et~al.(2012)]{Lugaz2012}
Lugaz, N. et al., 2012, 2012, Astrophys. J. 759, 68.

\bibitem[Lyot(1939)]{Lyot1939} Lyot, M.~B. 1939: Mon. Not. R. Astron.
Soc. 99, 578.

\bibitem[Lyot(1933)]{Lyot1933}Lyot, B. 1933: J. Roy. Astron. Soc.
Canada, 226, 265.

\bibitem[MacQueen(1993)]{MacQueen1993} MacQueen, R.~M., 1993: Solar
Phys. 145, 169.

\bibitem[MacQueen et~al.(1974)]{MacQueen1974} MacQueen, R.~M., Eddy,
J.~A., Gosling, J.~T., Hildner, E., Munro, R.~H., Newkirk, G.~A.,
Jr., Poland, A.~I., \&\ Ross, C.~L. 1974: Astrophys. J. 87, L85.

\bibitem[Minnaert(1930)]{Minnaert1930} Minnaert, M. 1930: Z. Astrophys.
1, 209.

\bibitem[Mizuno et~al.(2005)]{Mizuno2005}Mizuno, D.~R. et al. 2005:
J. Geophys. Res. 110, A07230.

\bibitem[Moran \&{} Davila(2004)]{Moran2004}Moran, T.~G. and Davila,
J.~M. 2004: Science 305, 66.

\bibitem[Moran et~al.(2010)]{Moran2010} Moran, T.~G., Davila, J.~M.,
\&\ Thompson, W.~T. 2010: Astrophys. J. 712, 453.

\bibitem[M\"ostl et~al.(2012)]{Mostl2012}
M\"ostl, C. et~al., 2012, Astrophys. J. 758, 10.

\bibitem[Poland \&\ Munro(1976)]{Poland1976} Poland, A.~I., \&\ Munro,
R.~H. 1976: Astrophys. J. 209, 927.

\bibitem[Richter et~al.(1982a)]{Richter1982a} 
Richter, I., Leinert, C. and Planck, B. 1982: Astron. Astrophys. 110, 111.

\bibitem[Richter et~al.(1982b)]{Richter1982b}Richter,
I., Leinert, C., and Planck, B. 1982: Astron. Astrophys. 110, 115.

\bibitem[Rust et~al.(1979)]{Rust1979} Rust, D.~M. 1979: In ``Physics
of solar prominences'', Proc. Coll. Oslo 1978, p.252.

\bibitem[Schuster(1879)]{Schuster1879} Schuster, A. 1879: Mon. Not.
R. Astron. Soc. 40, 35.

\bibitem[Tappin et~al.(2004)]{Tappin2004} Tappin, S.~J., Buffington,
A., Cooke, M.~P., Eyles, C.~J., Hick, P.~P., Holladay, P.~E.,
Jackson, B.~V., Johnston, J.~C., Kuchar, T., Mizuno, D., Mozer,
J.~B., Price, S., Radick, R.~R., Simnett, G.~M., Sinclair, D.,
Waltham, N.~R., Webb, D.~F. 2006: Geophys. Res. Lett. 31, L02802,
doi:10.1029/2003GL018766.

\bibitem[Tappin \&\ Howard(2009)]{Tappin2009} Tappin, S.~J., \&\ Howard,
T.~A. 2009: Space Sci. Rev. 147, 55.

\bibitem[van de Hulst(1950)]{vandehulst1950} van de Hulst, H.~C.
1950: Bull Astron. Inst. Neth. 11, 135.

\bibitem[Vourlidas \&\ Howard(2006)]{Vourlidas2006} Vourlidas, A.,
\&\ Howard, R.~A. 2006: Astrophys. J. 642, 1216.

\bibitem[van Houten(1950)]{vanHouten1950} van Houten, C.~J. 
1950:  Bull. Astron. Inst. Neth. 11, 160.

\bibitem[Wagner(1982)]{Wagner1982} Wagner, W.~J. 
1982: Adv. Sp. Res. 2, 203.

\bibitem[Webb et~al.(1980)]{Webb1980} Webb, D.~F., Cheng, C.-C.,
Dulk, G.~A., Edberg, S.~J., Martin, S.~F., McKenna-Lawler, S.,
\&\ McLean, D.~J. 1980: In Sturrock, P. (ed.) ``Solar Flares: A
monograph from Skylab Solar Workshop II'', p.471.

\bibitem[Webb \& Jackson(1987)]{Webb1987}
Webb, D.~F. and Jackson, B.~V., Proc. Sol. Wind VI, 2, 267.

\bibitem[Webb \& Jackson(1995)]{Webb1995}
Webb, D.~F. and Jackson, B.~V., Proc. Sol. Wind VIII, 97.

\bibitem[Webb et~al.(2006)]{Webb2006} Webb, D.~F., Mizuno, D.~R.,
Buffington, A., Cooke, M.~P., Eyles, C.~J., Fry, C.~D., Gentile,
L.~C., Hick, P.~P., Holladay, P.~E., Howard, T.~A., Hewitt, J.~G.,
Jackson, B.~V., Johnston, J.~C., Kuchar, T.~A., Mozer, J.~B.,
Price, S., Radick, R.~R., Simnett, G.~M., \&\ Tappin, S.~J. 2006:
J. Geophys. Res. 111, A12101.

\bibitem[Wood \&\ Howard(2009)]{Wood2009} Wood, B.~E., \&\ Howard,
R.~A. 2009: Astrophys. J. 702, 901.

\end{thebibliography}
\end{document}